# Defining an intrinsic "stickiness" parameter of stock price returns

Naji Massad♣,♣♣ and Jørgen Vitting Andersen♣,♣♣

October 2019

♣ CNRS, ♣♣Centre d'Economie de la Sorbonne, Université Paris 1 Pantheon-Sorbonne, Maison des Sciences Economiques,106-112 Boulevard de l'Hôpital, 75647 Paris Cedex 13, France.

**We introduce a non-linear pricing model of individual stock returns that defines a "stickiness" parameter of the returns. The pricing model resembles the capital asset pricing model (CAPM) used in finance but has a non-linear component inspired from models of earth quake tectonic plate movements. The link to tectonic plate movements happens, since price movements of a given stock index is seen adding "stress" to its components of individual stock returns, in order to follow the index. How closely individual stocks follow the index's price movements, can then be used to define their "stickiness".**

Keywords: non-linear CAPM, "stickiness" of stock returns,

## 1 Introduction

When investors want to assess the price of a given stock, in order, for example, to do risk management or portfolio allocation, they often use the basic capital asset pricing model (CAPM) [1-3] to guide them of a proper price/risk. The CAPM introduces a linear relationship between the return of a given asset, and the risk of the market. The risk of the market, is, in the CAPM framework defined in terms of variances and co-variances between the return of the asset and the excess return of the market. This measure of risk is called the market beta. For a longer discussion of the model, as well as some of the empirical failings of the CAPM, see for example Fama and French (2004) [4].

It is hard to underestimate the importance that the CAPM has had for the financial industry as a tool for pricing assets. Graham and Harvey (2001) [5] found in a survey of 392 CFOs in companies throughout the U.S. and Canada, that 74% of respondents always, or almost always, use the CAPM in estimating the cost of equity capital. In another study [6] Bruner, Eades, Harris, and Higgins (1998) found that 85% of their 27 best-practice firms use the CAPM or a modified version of the CAPM. While the CAPM is popular tool for practitioners to use for pricing of assets, it has in parallel been extensively studied in academic research papers, introducing "a zoo of new factors" [7]. Given such a plethora of factors and the inevitable data mining, many of the discovered factors probably were deemed «significant» by pure chance [8]. Harvey, Liu and Zhu (2015) [8] made an extensive testing of at least 316 linear factors models to explain the cross-section of expected returns. They argue that it is a serious mistake to use the usual statistical significance cutoffs in asset pricing tests.

A natural extension of the original CAPM would be to take into account betas that are time varying. Time varying properties can be captured in models like the autoregressive conditionally heteroskedastic (ARCH) model of Engle [9] and the generalized version (GARCH) by Bollerslev [10]. But such a description within a linear framework has been shown to have its own problems, even when allowing for time varying coefficients. Ghysel [11] for example showed that models with dynamical varying betas have stability problems. Often they are not able to give a proper description of the dynamics of the betas. The study by Ghysel found that constant beta models in many cases still yield on average better predictions compared to models with time varying betas. In essence, the time varying CAPM models have a tendency to overstate the time variation and as a result produce beta risk that is too volatile and changing too rapidly. It highlights the general problem with parameter instability in factor models like the CAPM and the APT (Arbitrage Pricing Theory). Still many would argue that the static CAPM should be replaced by some form of conditional or non-linear version of the CAPM.

One argument for conditional or non-linear versions of the CAPM relates to business-cycles, since market volatility is tightly linked to business-cycles. Lettau and Ludvigson (2001) [12] and Petkova and Zhang (2005) [13] both suggest that stocks have different exposures to market risk during recessions and expansions. In similar vein, Granger and Silvapulle (2001) [14] study whether or not beta responds asymmetrically to good and bad news. They define three market scenarios: bad, usual and good, conditional on the quantiles of the market return. Using the Dow Jones index over 9 years, they found that, for 21 stocks the beta of the stock is higher when the market is in a bearish state compared to a bullish state, while for the remaining 9 stocks the opposite is true. Their results confirm the general view that the betas increase (decrease) when the market is in a bearish (bullish) state. Another example of extending models to a non-linear description of the prices was introduced by Chen et al. [15]. They introduced a multiple-regime threshold GARCH model, in order to capture asymmetric risk, by allowing the market beta to change discretely between regimes that are driven by market information. In the following we introduce a new threshold model. However, in our case the threshold is not at the level of the market, but is instead introduced as an intrinsic property of a given stock.

**2 Theoretical framework: linear factor models, the CAPM, the yard-stick model**

In the capital asset pricing model the stock returns are estimated through a linear formula:

$$E(R_i) = E(R_f) + \beta_{iM} [E(R_M) - E(R_f)] \quad \text{(CAPM)} \quad (1)$$

Here $E(R_M)$ is the expected return of the market M, and $E(R_i)$ is the expected return of asset i. $E(R_f)$ is the expected return of the interest free rate, which, since we are only considering daily returns, will be taken equal to zero in the following. $\beta_{iM}$ is given by the covariance of the return of asset i with the market return divided by the variance of the market return: $\beta_{iM} = \frac{Covar(R_i, R_M)}{Var(R_M)}$.

We don't have access to people's expectation, so in practice the beta's are estimated via time-series regression:

$$R_i(t) = \beta_{iM} R_M(t) \qquad (2)$$

Insert Figure 1 around here

**Figure 1: "Stickiness" of the stock price movements.** *Aggregate price movements of the stocks (illustrated by the bar) pull each stock differently depending on the "coupling" constant $\beta_i$ and the "stickiness" parameter $R_{C_i}$ of each stock. In the case illustrated, the "stickiness" thresholds $R_{C_1}$ and $R_{C_3}$ prevent stocks 1 and 3 from moving, whereas stock 2 overcomes its threshold $R_{C_2}$ and moves.*

In the following we will assume the stocks being influenced by the aggregate price movements of the other N-1 stocks in an index, see Fig. 1, which illustrates the idea that each stock is linked to the market through a spring, with a spring constant given by a combination of two factors $\gamma_{i,M-i} * \alpha_{i,M-i}$. The first factor $\gamma_{i,M-i}$ measures, like the $\beta_{iM}$ of the CAPM, the specific sensitivity of the asset i's return to variation in the market return of the remaining N-1 assets. The second factor, $\alpha_{i,M-i}$ in addition weights the importance of the capitalization of a stock for the aggregate price movements, see Eq.(3) below. The stock price (represented by a small square) in this picture moves because of the aggregate price movement of the market represented by the "rod". In order to exclude any self-impact, we consider that the price of stock i moves because of the price movement of the N-1 other stocks constituting the index, with the return of the N-1 stocks, $R_{M-i}(t) \equiv \ln(\frac{\frac{1}{N-1}\sum_{j\neq i}^{N} S_j(t)}{\frac{1}{N-1}\sum_{j\neq i}^{N} S_j(t-1)})$. Here $S_j(t)$ is the price of stock j at time t (assuming for simplicity an index with equal weights on its assets).

$$R_i(t) = \alpha_{i,M-i} \gamma_{i,M-i} R_{M-i}(t) \quad ; \quad \alpha_{i,M-i} = [\, 1 - e^{\frac{-K_i}{K_{M-i}*\delta}} \,] \quad (3)$$

The constant $\alpha_{i,M-i}$ describes how a given stock moves in response to the market, depending on the ratio of capitalization of the given stock, $K_i$, to the capitalization of the remaining stocks in the index $K_{M-i} \equiv \sum_{j\neq i}^{N} K_j$. $\delta$ describes the scale of impact that the capitalization $K_{M-i}$ of the remaining stocks have on the price movement of stock i. A small $\delta \ll 1$ means that the price movement of stock i is dominated by the movement of the remaining stocks. The case $\delta \gg 1$ would correspond to a case of stocks moving independently, something we know empirically is not true, so a priori one should expect small values of $\delta$, but how small would define the proper impact of the remaining stocks on the price movement of stock i.

Insert Figure 2 around here

**Figure 2: Performance of linear pricing models.** *The plot a) shows the CAPM hypothesis Eq. (1) using the open-close returns of the Dow Jones Industrial index over the period 14/10/2016 to 10/03/2017. Plot b) illustrates instead Eq. (3) and c) Eq. (4) using same data set. Each point correspond to a daily open-close return $R_i$ of a given stock i.*

In order to test the CAPM on daily open-close returns, and the simple one factor model of Eq.(3) we show in Fig. 2a the CAPM hypothesis Eq. (2) using the daily open-close returns of the Dow Jones Industrial index over the period 14/10/2016 to 10/03/2017. Fig. 2b illustrates instead Eq. (3) using same data set. Each point corresponds to a daily return of a given stock i (x-axis), and its prediction (y-axis). The data points show the results of out-of-sample simulations using first an in-sample period (from 23/05/2016 to 13/10/2016) to determine the $\alpha_{iM-i}$'s of Eq.(2) and the $\gamma_{i,M-i}$, $\delta$ of Eq.(3). For a given fixed value of $\delta$, the set of $\gamma_{i,M-i}$'s that minimizes the error of forecasting of $R_i$ can be obtained directly from Eq.(3). The optimal value of delta was found in-sample to be very small, as expected, $\delta^{optimal} = 0.00102$. In order to smooth out the fluctuations of $\gamma_{i,M-i}$, we used aggregate returns over a large window of size T=250. As seen from Fig. 2b the mean error in predicting $R_i$ is larger compared to the CAPM case, but the skewness and kurtosis of the error is smaller compared to the CAPM case.

In [16] we have instead pointed out the importance of a relative sentiment measure of a given stock to its peers. The idea is that people use heuristics, or "rules of thumb", in terms of "yard sticks" from the performance of the other stocks in a stock index. The under-/over-performance with respect to a yard stick happens because of a general negative/positive sentiment of the market participants towards a given stock. The bias created in such cases does not necessarily have a psychological origin but could be due to insider information. Insiders having superior information about the state of a company reduce/increase their stock holding gradually causing a persisting bias over time. The introduction of a measure for the relative sentiment of a stock has allowed us to come up with another one factor model, very similar in structure to the CAPM model [16]:

$$R_i(t) = \gamma_{iM-i} R_{M-i}(t) \qquad \gamma_{iM-i} = \frac{\sigma(R_i)}{\sigma(R_{m-i})} \quad ; \quad R_{M-i}(t) \equiv \ln\left(\frac{\frac{1}{N-1}\sum_{j\neq i}^{N} S_j(t)}{\frac{1}{N-1}\sum_{j\neq i}^{N} S_j(t-1)}\right),$$

(4)

In Eq.(4) $\gamma_{iM-i}$ now describes the relative volatility of the stock to the volatility of the market of the remaining N-1 stocks. Therefore the basic pricing mechanism via Eq.(4) is given by traders using the Sharpe ratio of a stock and comparing it to the Sharpe ratio of the market as a yard-stick for whether a stock is under/overvalued. Figure 2b shows daily performances of the stocks to have a clear tendency to cluster around the measures introduced by the yard sticks in accordance with the pricing formula Eq.(4). However, the error of the prediction of $R_i$ has a larger mean and higher order moments larger compared to the CAPM case. In other out-of-sample intervals we found larger error and standard deviations, but smaller skewness and kurtosis compared to the CAPM case.

The CAPM, Eq.(3), and Eq.(4) presented above are examples of one-factor models which despite their simplicity often are able to capture quite specific behavior. Cizeau et al. [17] showed how correlation between stock returns in extreme conditions could be satisfactorily described by time independent factors without having to invoke correlations that changes over time. It is important to note a qualitative property of the factor models presented above, namely that although the market is built from the fluctuations of the stocks, it is the market that is the fundamental quantity, not the stocks themselves. Therefore, you can't explain statistical properties of the market returns from those of the stocks returns within such models. In that respect we would like to point out the difference here between stocks fluctuating because they are coupled to an index, and fluctuations across markets worldwide. In the case of spillover effects across markets, it *does* make sense to believe that the rise of correlations is caused by increasing fear by investors, something also pointed out by Balogh et al. [18]. Their argument for correlations across markets happens via prospect theory [19], a very different mechanism compared to stocks following almost automatically a given index.

**3 Theoretical framework: a non-linear stick-slip factor model,**

We now suggest to go beyond a linear description by introducing a "stickiness" parameter of individual stock returns. As illustrated in Fig.1, a small price movement of the market (illustrated by the moving rod) does not necessarily induce a price movement of all individual stocks prices, e.g. in Fig. 1 the stock 1 and 3 don't move (they "stick") whereas stock 2 follows the market price movement (it "slips"). As mentioned in [20], a psychological reason why the small price movements get ignored could be because of the so-called "change blindness" [21], a behavioral trait which reflects the tendency of humans to reply in a nonlinear fashion to changes. This is consistent with experiments made in psychology which have shown that humans react disproportionally to big changes, whereas small changes go unnoticed [21–24]. Another reason for a non-linear reaction of market participants could be insider knowledge: people with knowledge of negative news for a stock, could persistently use positive market movements as a selling opportunity, creating inertia in individual stock price movements. To take into account such effects we suggest the following equations:

$$R_i(t) = \left(\frac{\sigma(R_i)}{\sigma(R_{m-i})}\right) \times \Theta\left(\left[\frac{\sigma(R_i)}{\sigma(R_{m-i})} |R^{cum}_{m-i}(t)| > R_{C_i}\right]\right) \times R^{cum}_{m-i}(t)$$

(5)

$$R^{cum}_{m-i}(t) = (1-\Theta\left(\left[\frac{\sigma(R_i)}{\sigma(R_{m-i})} |R^{cum}_{m-i}(t-1)| > R_{C_i}\right]\right)) \times R^{cum}_{m-i}(t-1) + R_{m-i}(t)$$

(6)

Eq.5 describes the "change blindness" part since the stock i only gets a contribution from the remaining N-1 stocks when the normalized cumulative return (or "stress"), $\frac{\sigma(R_i)}{\sigma(R_{m-i})}|R_{m-i}^{cum}(t)|$, exceeds a certain threshold $R_{C_i}$ which is specific to stock i. Eq.(6) describes how the "stress" on stock i, from the remaining N-1 stocks, keeps on cumulating until it exceeds $R_{C_i}$, after which it is reset to zero: the effect of the price movements of the N-1 stocks becomes "prized in". For a very small $R_{C_i}$ the condition in the Theta function is always fulfilled, so the model in this case coincide with the yard-stick model Eq.4. As stressed before, this case corresponds to traders comparing the Sharpe ratio of a stock to the Sharpe ratio of the market, in order to decide whether a stock is over/under-valued. A large value $R_{C_i}$ instead means that the price of stock i only changes when the aggregate price movements of the market exceeds a certain threshold, determined by the Sharpe ratio of the aggregate market movement and a threshold Sharpe ratio of stock i given by $R_{C_i}/\sigma(R_i)$. As seen from Eq.6 after a price change has happened at say time t-1, $R_{m-i}^{cum}$ is reset to zero (the market price movement has become "priced-in") before another market price movement eventually again adds to $R_{m-i}^{cum}$ at time t. Therefore one can consider the variable $R_{m-i}^{cum}(t)$ as an oscillator: consecutive market movements keeps on adding to $R_{m-i}^{cum}$ up to the value (positive or negative) determined by the threshold condition in the Theta function of Eq.5,6, after which $R_{m-i}^{cum}$ is set to zero and another oscillation can begin. The period of the oscillation is determined by the fraction of the Sharpe ratio of the aggregate market movement, and the threshold Sharpe ratio of stock i, given by the $R_{C_i}/\sigma(R_i)$. Fig. 3 illustrate such oscillations for three different values of $R_{C_i}$ for stocks of the Dow Jones index.

Insert Figure 3 around here

**Figure 3: Oscillations of the "stress" field $R_{m-i}^{cum}(t)$ for three different stickiness values $R_{C_i}$.**

*Top plots show the cumulative market return (red line), stock return (blue line) for the stock American Express, and prediction of the stock return Eq.(5)-(6) (green line) over the time period 10/04/2015-07/07/2015. The plots a)-c) correspond to three different values of the stickiness parameter $R_{C_i}$. The bottom plots show the oscillations of the "stress" field $R_{m-i}^{cum}(t)$.*

The pricing model resembles the capital asset price model (CAPM) used in finance but has a non-linear component with origin from models of earth quake tectonic plate movements [25]. The link to tectonic plate movements happens, since price movements of a given stock index is seen adding "stress" to its components of individual stock returns, in order to follow the index. How closely individual stocks follow the index's price movements, can then be used to define its "stickiness".

**4 Results: defining intrinsic "stickiness" of stocks**

In order to test the model, we first applied equations Eqs.(5)-(6) to the Dow Jones Industrial Index over the period 23/03/2015 to 19/04/2016. The initial choice of this special period was because the composition of the Dow did not change during that period, other periods were subsequently chosen to test for stationarity of the results. Since we are studying daily data, and since the trading volume between the opening and the close by far exceeds the trading volume between the close and the open, we only consider the daily return between the open and the close [16]. For each stock of the Dow Jones, we applied the discrete equations Eqs.(5)-(6). The ratio $\frac{\sigma(R_i)}{\sigma(R_{m-i})}$ was calculated using a running window of 20 days. In order to determine the "stickiness" parameter, $R_{C_i}$, of each stock i, we minimized the error function, $E_i$:

$$E_i(R_{C_i}) = \sum_t (R_i^{model}(t) - R_i^{data}(t))^2$$

(7)

using Eqs.(5)-(6) to obtain $R_i^{model}(t)$.

Insert Figure 4 around here

**Figure 4: Bootstrap to define confidence levels of the stickiness parameter.** *Blue line shows the error of the model, Eq.(7), as a function of the "stickiness" parameter, $R_{C_i}$, for the Boing Co. stock over the time period 23/03/2015-19/04/2016. The red line in figure 4 represents the level of the 10th percentile, whereas the green line represents the level of 90th percentile using in total 100 re-shuffled data samples.*

The blue line in Fig. 4 shows the error of the model, Eq.(7), as a function of the "stickiness" parameter, $R_{C_i}$ for one given stock, the Boing Co. stock. As can be seen, for this particular stock, for small values of $R_{C_i}$ the prediction of $R_i(t)$ using Eq.5-6 does not seem very different from the yard-stick model Eq.(2), which corresponds to the case where $R_{C_i} \equiv 0$. For larger values of $R_{C_i}$ a global minimum, $R_{C_i}^{min} \approx 0,0052$, of the error function $E_i$ appear before the error finally increases for large values of $R_{C_i}$. It is easy to understand the large prediction errors that happens for large values of $R_{C_i}$, by looking at figure 3. A large value of $R_{C_i}$ means that the stock only reacts for large cumulative price movements of the market, thereby "overshooting" the stock's observed price movement, when the condition in the Theta function in Eqs.(5)-(6) is finally met. We therefore use the observed minimum of $R_{C_i}$ as a measure of what we call intrinsic "stickiness" of a stock.

The question however is whether our finding of a $R_{C_i} > 0$ is due to the inherent noise of the time series. In order to determine this, we applied boots trapping. By reshuffling the original return price time series (excluding the first 20 days that serve to calibrate the volatility ratio), we break any temporal correlations (but still keep any overall trend of the market), that could contain information used to predict the price returns via our model. The red line in Fig. 4 represents the level of the 10$^{th}$ percentile, whereas the green line represents the level of 90$^{th}$ percentile using in total 100 samples. We checked that using instead 1000 samples we obtained the same qualitative behavior. As can be seen for the BA stock we were able to find a $R_{C_i}^{min}$ below the 10$^{th}$ noise level. We then applied this method to the remaining 29 stocks of the Dow index, see table 1. First column gives the noise level for the yard-stick model Eq.(4) which by definition corresponds to the $R_{C_i} \equiv 0$ case of Eqs.(5)-(6). Second column gives the level of the 10$^{th}$ percentile found from bootstrapping, third column gives $R_{C_i}^{min}$, and forth column then indicates the value of the "stickiness" $R_{C_i}^{min}$ for those stocks where we were able to define a such. In order to define a "stickiness" $R_{C_i}^{min}$ of a stock we required $E_i(R_{C_i}^{min}) < E_i(R_{C_i} \equiv 0)$ AND $E_i(R_{C_i}^{min}) < E_i^{10th}(R_{C_i})$ with $E_i()$ the error of stock i, and $E_i^{10th}$ denoting the noise level of the 10$^{th}$ percentile. That is, in order to claim to have found a "stickiness" $R_{C_i}^{min}$ of a stock i, it needs to give a better prediction compared to the case of the model without threshold, ($R_{C_i} \equiv 0$), as well as have a noise level that is below the 10$^{th}$ percentile found via the boots trap method.

**Table 1. 14 stocks have a Rc that is above 0 (23/03/2015-19/04/2016)**

| Stock | Value for Rc=0 | Noise level | Min error | Rc |
|---|---|---|---|---|
| AAPL | 0.0462 | 0.0459 | 0.0460 | - |
| AXP | 0.0336 | 0.0330 | 0.0330 | - |
| BA | 0.0293 | 0.0291 | 0.0288 | 0,00519 |
| CAT | 0.0449 | 0.0446 | 0.0448 | - |
| CSCO | 0.0178 | 0.0178 | 0.0178 | - |
| CVX | 0.0555 | 0.0548 | 0.0547 | 0,00247 |
| DIS | 0.0250 | 0.0244 | 0.0245 | - |
| DWDP | 0.0369 | 0.0367 | 0.0368 | - |
| GE | 0.0167 | 0.0165 | 0.0161 | 0,00346 |
| GS | 0.0268 | 0.0261 | 0.0266 | - |
| HD | 0.0227 | 0.0218 | 0.0215 | 0,00432 |
| IBM | 0.0225 | 0.0221 | 0.0222 | - |
| INTC | 0.0289 | 0.0288 | 0.0289 | - |
| JNJ | 0.0119 | 0.0117 | 0.0117 | 0,00227 |
| JPM | 0.0139 | 0.0138 | 0.0138 | 0,00172 |
| KO | 0.0127 | 0.0125 | 0.0125 | 0,00213 |
| MCD | 0.0186 | 0.0181 | 0.0181 | 0,00255 |
| MMM | 0.0115 | 0.0114 | 0.0114 | - |
| MRK | 0.0276 | 0.0272 | 0.0274 | - |
| MSFT | 0.0317 | 0.0311 | 0.0313 | - |
| NKE | 0.0432 | 0.0419 | 0.0415 | 0,0048 |
| PFE | 0.0296 | 0.0292 | 0.0293 | - |
| PG | 0.0153 | 0.0149 | 0.0152 | - |
| TRV | 0.0132 | 0.0129 | 0.0128 | 0,0028 |
| UNH | 0.0502 | 0.0494 | 0.0494 | 0,00715 |
| UTX | 0.0159 | 0.0157 | 0.0155 | 0,00424 |
| V | 0.0302 | 0.0297 | 0.0294 | 0,00476 |
| VZ | 0.0131 | 0.0130 | 0.0129 | 0,00451 |
| WMT | 0.0415 | 0.0408 | 0.0413 | - |
| XOM | 0.0349 | 0.0348 | 0.0348 | - |

Using this definition of stickiness, as can be seen from table 1 we are then able to define 14 stocks with a stickiness in the period 23/03/2015-19/04/2016. As expected we only find small values of the stickiness parameter $R_{c_i}^{min}$, ranging from 0.17% for J.P. Morgan Chase up to 0.72% for UnitedHealth Group Inc. In order to test for stationarity, we show in the next two tables the results for the time periods 22/03/2016-18/04/2017 (table 2) as well as 21/03/2017 to 12/03/2018 (table 3). It should come as no surprise that we find the stickiness (as defined above) changes over time, just as example given correlations changes over time.

**Table 2. 8 stocks have a Rc that is above 0 (22/03/2016-18/04/2017)**

| Stock | Value for Rc=0 | Noise level | Min error | Rc |
|---|---|---|---|---|
| AAPL | 0.0216 | 0.0208 | 0.0205 | 0.0058 |
| AXP | 0.0193 | 0.0191 | 0.0191 | 0.00213 |
| BA | 0.0290 | 0.0288 | 0.0289 | - |
| CAT | 0.0373 | 0.0366 | 0.0370 | - |
| CSCO | 0.0126 | 0.0121 | 0.0122 | - |
| CVX | 0.0215 | 0.0211 | 0.0213 | - |
| DIS | 0.0118 | 0.0111 | 0.0118 | - |
| DWDP | 0.0300 | 0.0297 | 0.0296 | 0.00215 |
| GE | 0.0129 | 0.0126 | 0.0129 | - |
| GS | 0.0322 | 0.0321 | 0.0322 | - |
| HD | 0.0190 | 0.0185 | 0.0184 | 0.00588 |
| IBM | 0.0152 | 0.0150 | 0.0148 | 0.00382 |
| INTC | 0.0204 | 0.0197 | 0.0197 | - |
| JNJ | 0.0145 | 0.0141 | 0.0142 | - |
| JPM | 0.0146 | 0.0144 | 0.0145 | - |
| KO | 0.0153 | 0.0149 | 0.0152 | - |
| MCD | 0.0176 | 0.0168 | 0.0170 | - |
| MMM | 0.0074 | 0.0072 | 0.0072 | - |
| MRK | 0.0464 | 0.0460 | 0.0460 | 0.00426 |
| MSFT | 0.0155 | 0.0153 | 0.0155 | - |
| NKE | 0.0415 | 0.0405 | 0.0399 | 0.00633 |
| PFE | 0.0249 | 0.0242 | 0.0246 | - |
| PG | 0.0163 | 0.0154 | 0.0160 | - |
| TRV | 0.0165 | 0.0161 | 0.0165 | - |
| UNH | 0.0265 | 0.0256 | 0.0261 | - |
| UTX | 0.0097 | 0.0096 | 0.0097 | - |
| V | 0.0222 | 0.0216 | 0.0220 | - |
| VZ | 0.0234 | 0.0230 | 0.0226 | 0.0082 |
| WMT | 0.0279 | 0.0264 | 0.0269 | - |
| XOM | 0.0205 | 0.0201 | 0.0205 | - |

**Table 3. 5 stocks have a Rc that is above 0 (21/03/2017 to 12/03/2018)**

| Stock | Value for Rc=0 | Noise level | Min error | Rc |
|---|---|---|---|---|
| AAPL | 0.0360 | 0.0339 | 0.0355 | - |
| AXP | 0.0150 | 0.0148 | 0.0149 | - |
| BA | 0.0461 | 0.0444 | 0.0455 | - |
| CAT | 0.0294 | 0.0282 | 0.0291 | - |
| CSCO | 0.0181 | 0.0177 | 0.0180 | - |
| CVX | 0.0259 | 0.0251 | 0.0254 | - |
| DIS | 0.0412 | 0.0393 | 0.0408 | - |
| DWDP | 0.0302 | 0.0286 | 0.0297 | - |
| GE | 0.1070 | 0.1060 | 0.1048 | 0.0097 |
| GS | 0.0383 | 0.0375 | 0.0381 | - |
| HD | 0.0289 | 0.0282 | 0.0288 | - |
| IBM | 0.0165 | 0.0158 | 0.0159 | - |
| INTC | 0.0390 | 0.0372 | 0.0383 | - |
| JNJ | 0.0235 | 0.0231 | 0.0234 | - |
| JPM | 0.0180 | 0.0176 | 0.0175 | 0.0036 |
| KO | 0.0173 | 0.0171 | 0.0172 | - |
| MCD | 0.0211 | 0.0203 | 0.0209 | - |
| MMM | 0.0136 | 0.0133 | 0.0135 | - |
| MRK | 0.0250 | 0.0247 | 0.0249 | - |
| MSFT | 0.0210 | 0.0202 | 0.0202 | - |
| NKE | 0.0388 | 0.0369 | 0.0372 | - |
| PFE | 0.0192 | 0.0189 | 0.0188 | 0.00154 |
| PG | 0.0149 | 0.0145 | 0.0148 | - |
| TRV | 0.0272 | 0.0266 | 0.0263 | 0.00617 |
| UNH | 0.0266 | 0.0256 | 0.0266 | - |
| UTX | 0.0242 | 0.0236 | 0.0242 | - |
| V | 0.0168 | 0.0163 | 0.0168 | - |
| VZ | 0.0328 | 0.0323 | 0.0322 | 0.00304 |
| WMT | 0.0373 | 0.0367 | 0.0369 | - |
| XOM | 0.0156 | 0.0153 | 0.0155 | - |

In the second time period, 22/03/2016-18/04/2017, (table 2) we are only able to identify 8 stocks with a stickiness parameter $R_{c_i}^{min}$>0. Out of the 8 stocks, 3 were also identified having a stickiness in the first time period. In the third time period, 21/03/2017-12/03/2018 (table 3) we find only 5 stocks having a stickiness parameter $R_{c_i}^{min}$>0. However out of the 5 stocks, 4 were found also to have a stickiness parameter in the first time period. We find one stock with a stickiness parameter over all three time periods, Verizon Communications Inc. A priori it would seem reasonable to expect internet related companies, to be more "detached" from the general price movements of the index, just as we found for the Verizon stock. However, the fact that we are not able (over the given time periods) to define a stickiness parameter for neither IBM, nor Cisco, seems to contradict the results found for Verizon. Finally, we tried the method in the case where one replace the ratio $\frac{\sigma(R_i)}{\sigma(R_{m-i})}$ by $\frac{Covar(R_i,R_M)}{Var(R_M)}$, in Eq.5-6. In this case $R_{C_i} \equiv 0$, then coincides with the CAPM model. In general, several overlaps were found in the stocks for which we could define a stickiness parameter, so whether using the covariance, or just simple ratios of the standard deviations, seem less relevant when it comes to define a stickiness parameter of a given stock.

## 5 Discussion

We have introduced and studied two linear pricing models of individual stock returns and compared with the well-known capital asset price model (CAPM) in finance. After an initial discussion on linear pricing models, we then extended our study and considered a non-linear pricing model of individual stock returns that enables us to define a "stickiness" parameter of the returns. The pricing model resembles the CAPM, but has a non-linear component inspired from models of earth quake tectonic plate movements. The link to tectonic plate movements happens, since price movements of a given stock index is seen adding "stresses" to its components of individual stock returns, in order to follow the index. How closely individual stocks follow the index's price movements, can then be used to define their "stickiness". Using a bootstrap method, we attributed a statistical significance measure in order to define a stickiness parameter for each individual stock. In this way we were able to attribute a stickiness parameter for a number of stocks in the Dow Jones Industrial average. Our hope is that the introduction of stickiness allow for an enlarged understanding in the pricing process of stocks. In practical terms, one could for example imagine it be used as a tool in connection with the detection of insider trading, where insider knowledge introduces "friction" in price formation which could be captured in terms of our stickiness measure.

Acknowledgments

This work was carried out in the context of the Laboratory of Excellence on Financial Regulation (Labex ReFi) supported by PRES heSam under the reference ANR-10-LABX-0095. It benefitted from French government support managed by the National Research Agency (ANR) as part of the project Investissements d'Avenir Paris Nouveaux Mondes (invesiments for the future, Paris-New Worlds) under the reference ANR-11-IDEX-0006-02.

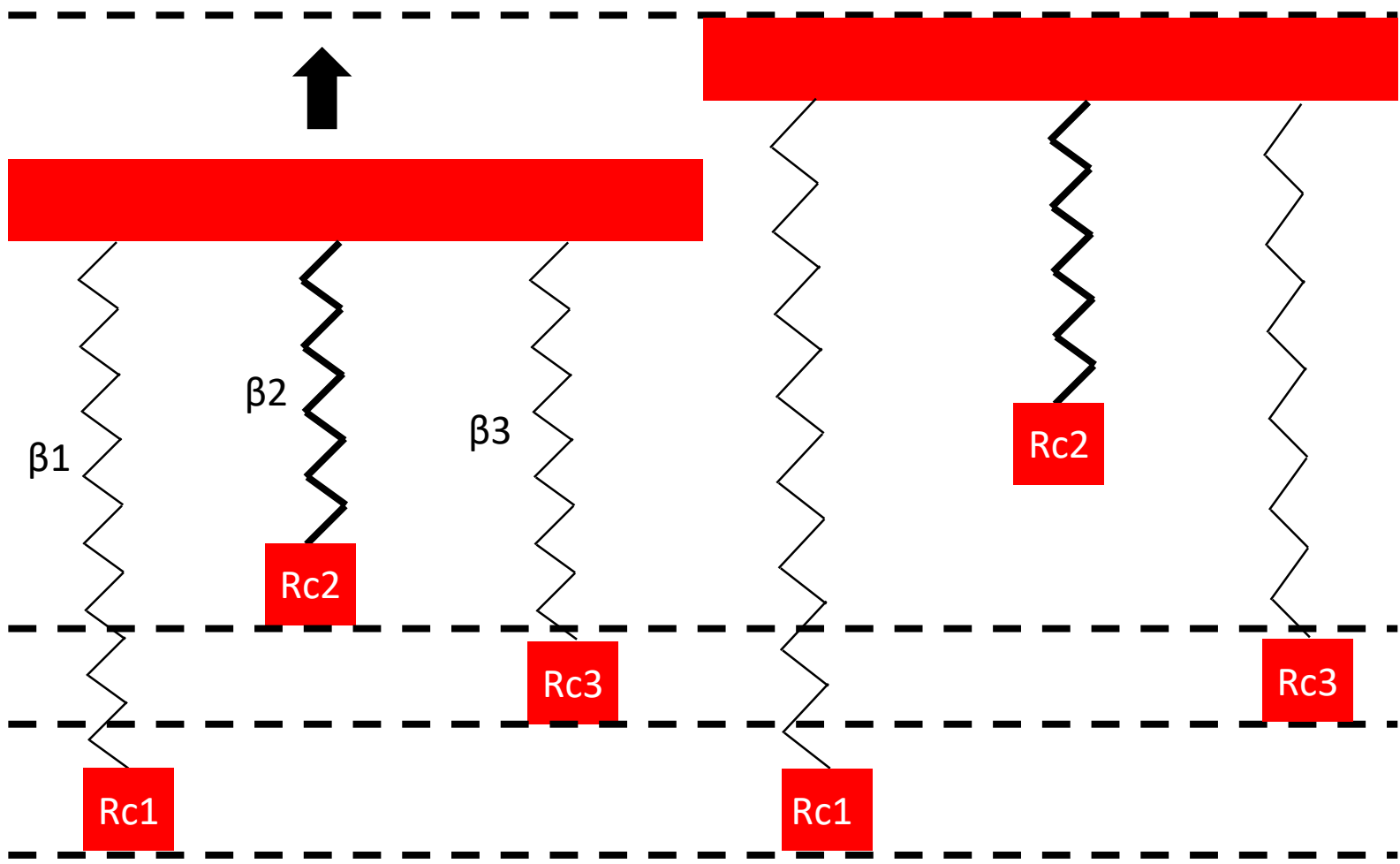

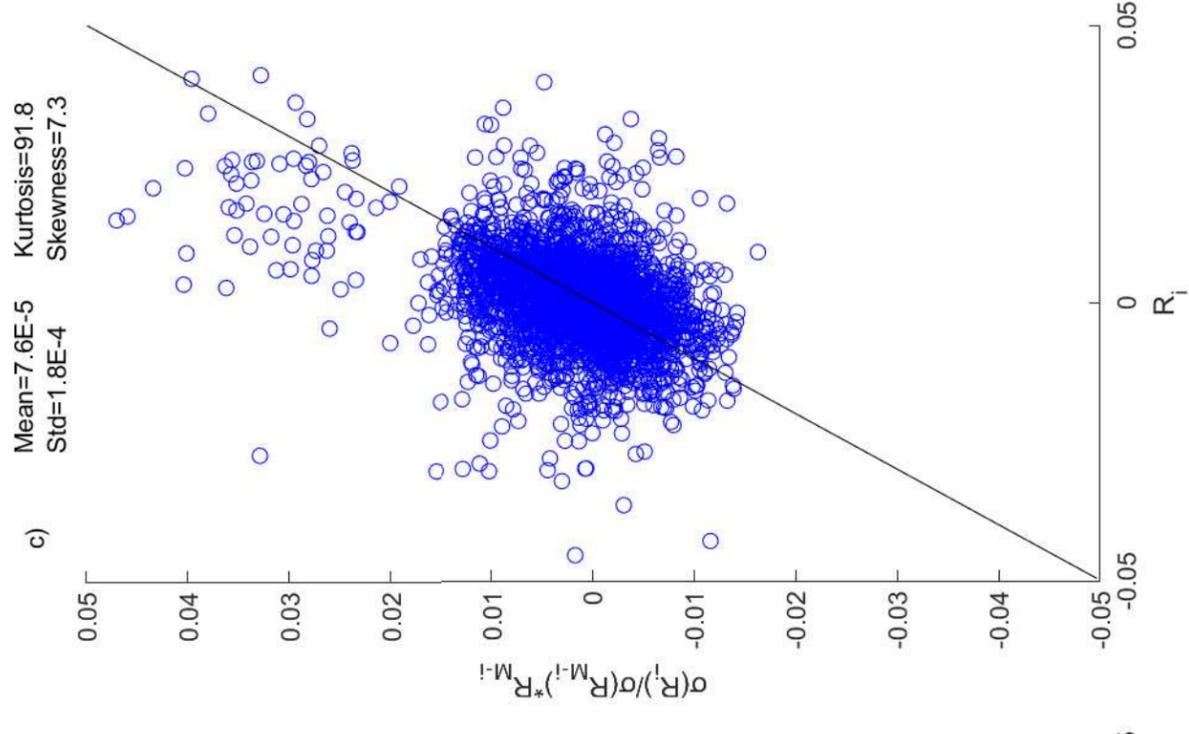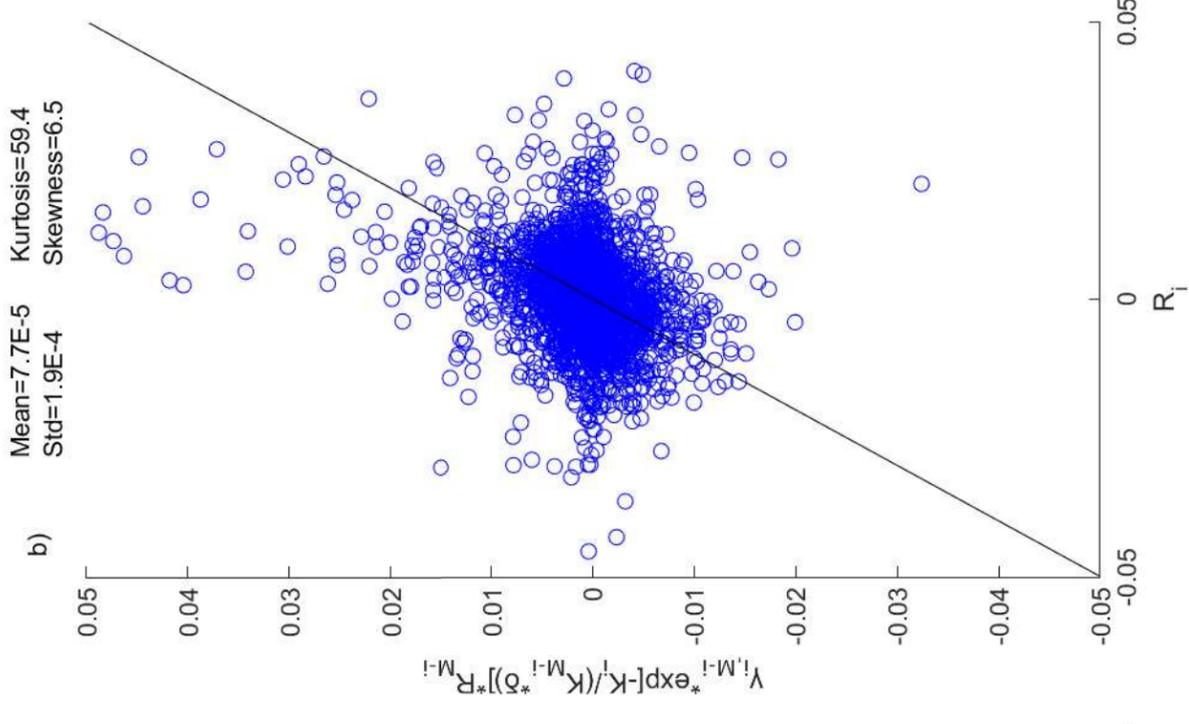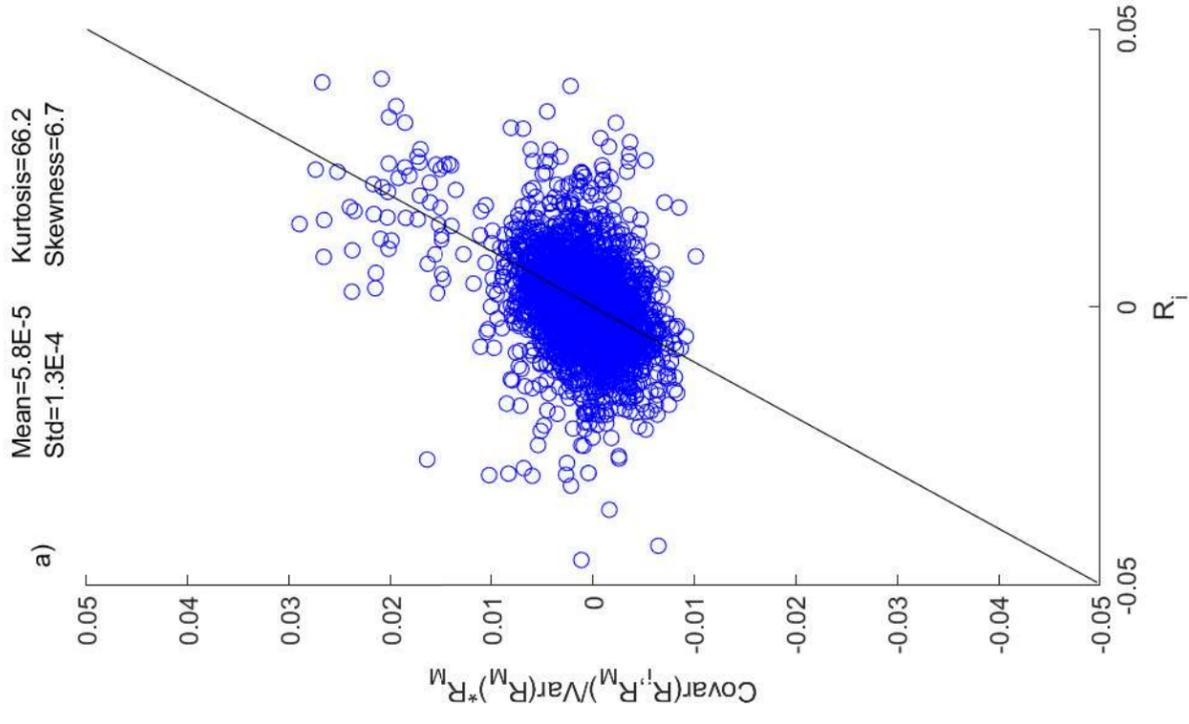

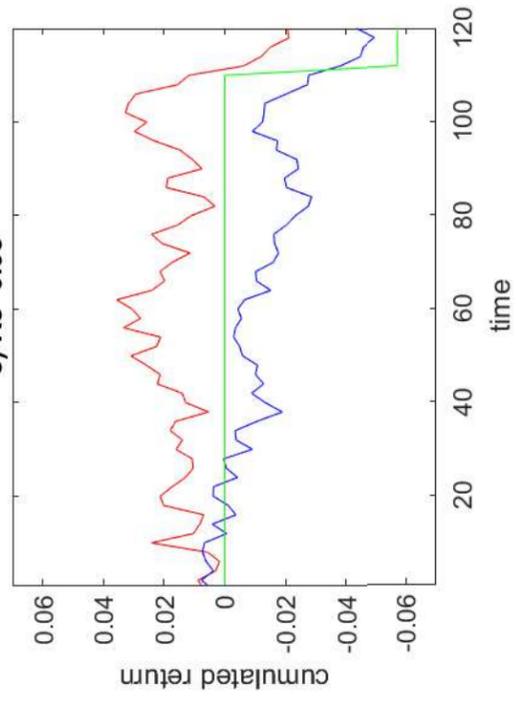
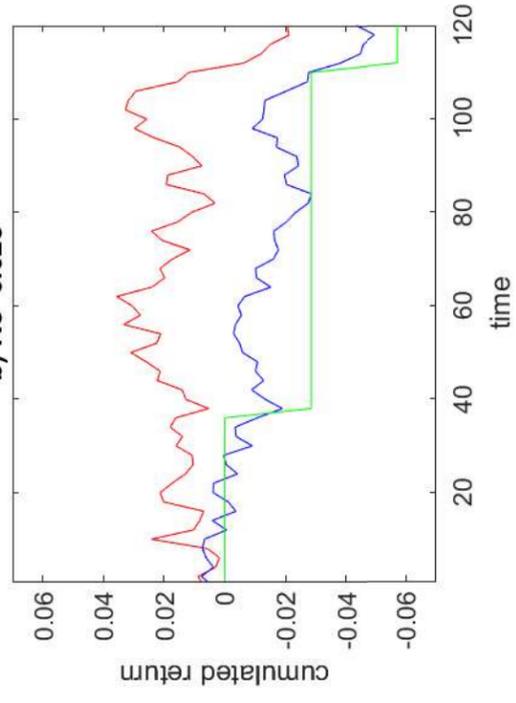
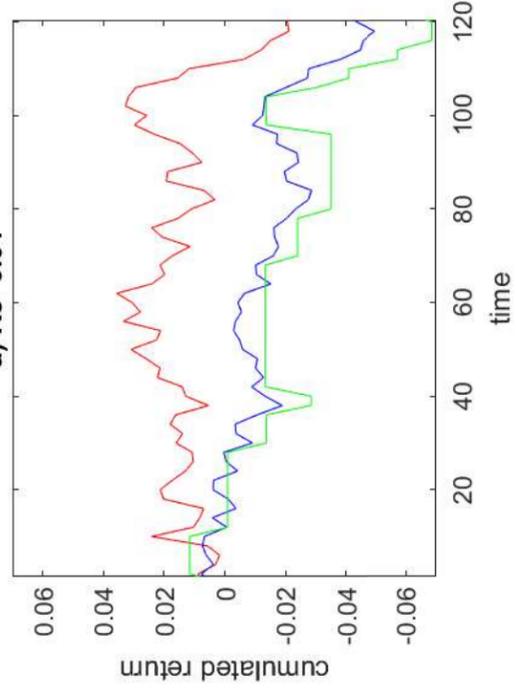
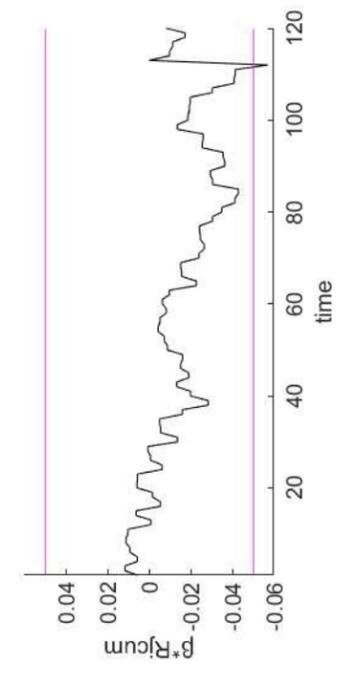
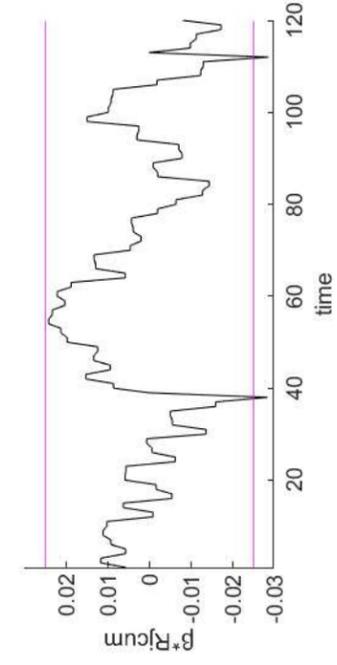
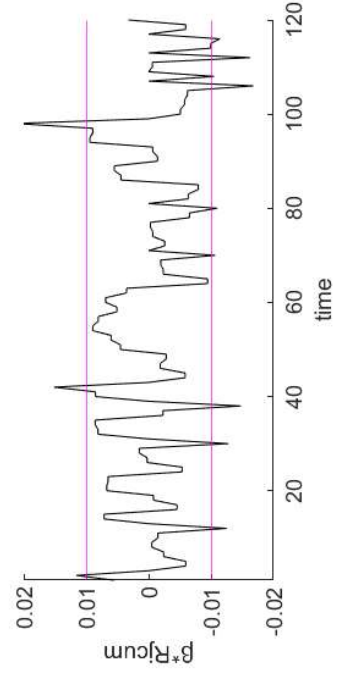

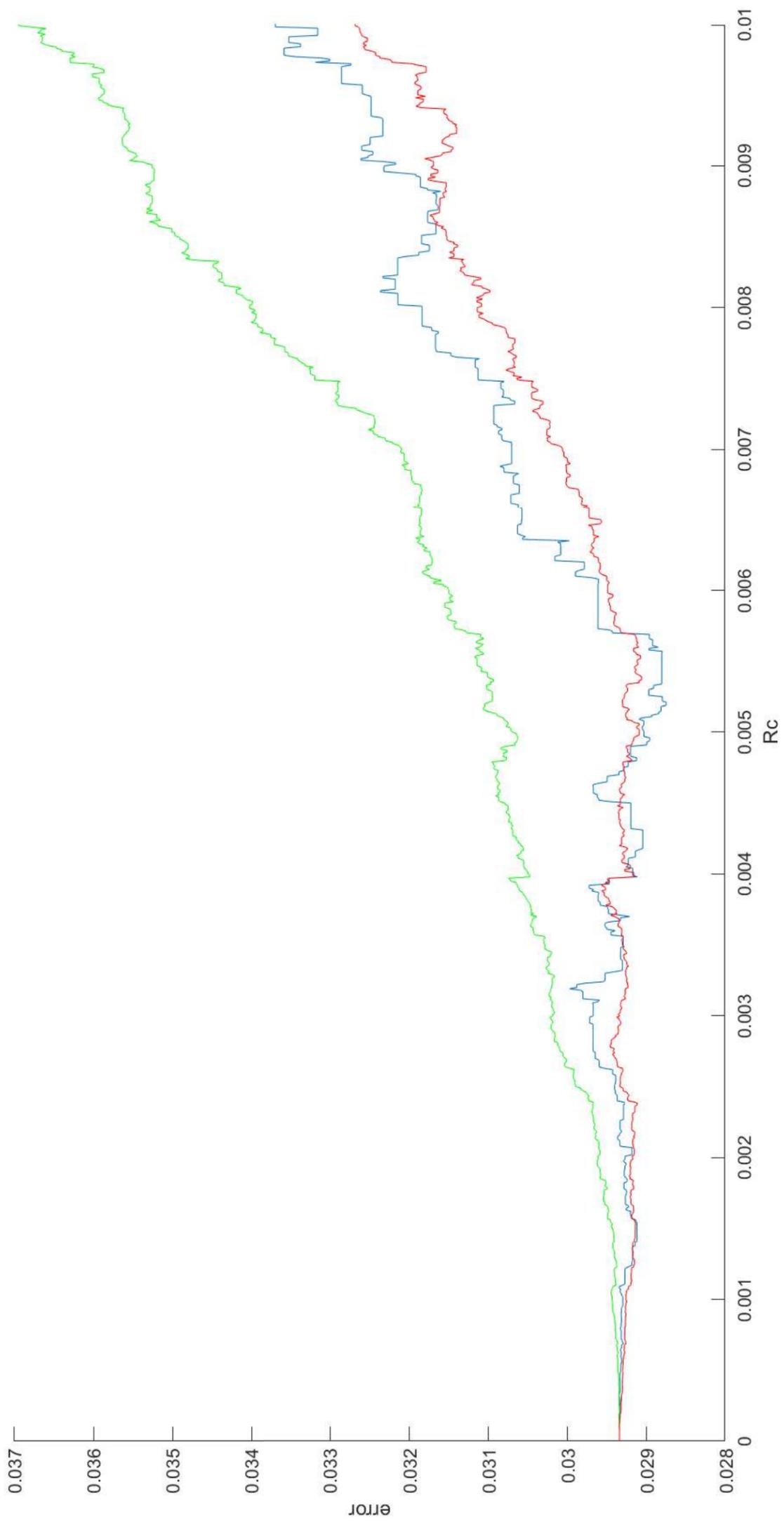